\documentclass[twocolumn]{aastex631}
\usepackage{xspace}
\usepackage{amsmath}
\usepackage{graphicx}

\newcommand{\hmpc}{h^{-1}{\rm Mpc}}
\newcommand{\hkpc}{h^{-1}{\rm kpc}}
\newcommand{\ihmpc}{\,h{\rm Mpc}^{-1}}
\newcommand{\hgpc}{h^{-1}{\rm Gpc}}

\newcommand{\planck}{\textit{Planck}\xspace}
\shorttitle{}
\shortauthors{}

\graphicspath{{./}{figures/}}

\begin{document}

\title{Signatures of Light Massive Relics on nonlinear structure formation}

\author[0000-0002-5209-1173]{Arka Banerjee}
\affiliation{Fermi National Accelerator Laboratory, Cosmic Physics Center, Batavia, IL 60510, USA \\}

\author{Subinoy Das  }
\affiliation{Indian Institute of Astrophysics, Bengaluru, Karnataka 560034, India}

\author{Anshuman Maharana}
\affiliation{Harish-Chandra Research Institute, A CI of  Homi Bhabha National Institute, Chhatnag Road, Jhunsi, Allahabad, UP 211019, India}

\author{Ravi kumar sharma}
\affiliation{Indian Institute of Astrophysics, Bengaluru, Karnataka 560034, India}

\begin{abstract}
Cosmologies with Light Massive Relics (LiMRs) as a subdominant component of the dark sector are well-motivated from a particle physics perspective, and can also have implications for the $\sigma_8$ tension between early and late time probes of matter clustering. The effects of LiMRs on the Cosmic Microwave Background (CMB) and structure formation on large (linear) scales have been investigated extensively. In this paper, we initiate a systematic study of the effects of LiMRs on smaller, nonlinear scales using cosmological $N$-body simulations; focusing on quantities relevant for photometric galaxy surveys. For most of our study, we use a particular model of nonthermal LiMRs but the methods developed easily generalize to a large class of models of LiMRs --- we explicitly demonstrate this by considering the Dodelson-Widrow form of the velocity distribution. We find that, in general, the effects of LiMR on small scales are distinct from those of a $\Lambda$CDM universe, even when the value of $\sigma_8$ is matched between the models. We show that weak lensing measurements around massive clusters, between $\sim 0.1\hmpc$ and $\sim 10 \hmpc$, should have sufficient signal-to-noise in future surveys to distinguish between $\Lambda$CDM and LiMR models that are tuned to fit both CMB data and large (linear) scale structure data at late times. Furthermore, we find that different
LiMR cosmologies which are indistinguishable by conventional linear probes can be distinguished by these probes if their velocity distributions are sufficiently different. LiMR models can, therefore, be best tested and constrained by jointly analyzing data from CMB and late-time structure formation on both large \textit{and} small scales.

\end{abstract}


\section{Introduction} \label{sec:intro}

   The search for the microphysical constituents of Dark Matter, or the dark sector, which interacts feebly with the visible sector, continues to drive research at the intersection of particle physics and cosmology. While much of the early focus of Dark Matter model building was in the Weakly Interacting Massive Particle (WIMP) paradigm, with one effectively cold and collisionless component, in recent years, multiple well-motivated models have been proposed that posit the existence of a more complicated dark sector.

A subset of such models predict the existence of light but massive particles, which constitute a subdominant component of the dark sector, and therefore, of the total energy budget of the Universe \citep[e.g.][]{2007JHEP...03..120C,2007PhRvD..75k5001F,2008JHEP...07..124A,Burgess:2008ri,2008JHEP...12..104A,2009PhRvD..80a5003E,2010PhRvD..81l3530A,2013PhRvD..88a5032B,2015PhRvD..92e5033C,2016PhRvL.117y1801A}. In some extended neutrino sector theories  beyond the standard model of particle physics,  light massive relics can easily have mass in \rm{eV}  range  \citep{ Das:2006ht,Abellan:2021bpx,Bjaelde:2010vt}. Other hidden sector relics with different particle physics origins are also discussed in, e.g. \citet{Das:2010ts,Ko:2017uyb,Bogorad:2021uew} . These Light Massive Relics (LiMRs in the rest of the paper) can have effects at both the epoch of the Cosmic Microwave Background (CMB), as well as in the late-time Universe. If the particles are not fully nonrelativistic at the time of recombination, they effectively act as ``dark radiation", and their effects can be captured by looking for deviations in $N_{\rm eff}$ from that expected in a Universe with no additional light degrees of freedom, apart from those in the Standard Model. In the late Universe, the LiMRs can become non-relativistic, but their velocity dispersion can still be high compared to peculiar velocities sourced by gravitational evolution. This leads to an effective free-streaming scale, below which perturbations of the LiMR component are damped. If the LiMRs make up a non-negligible fraction of the energy budget, this leads to an overall suppression of the growth of matter perturbation on small scales. In recent years, multiple studies have studied constraints on such models by combining early Universe, e.g the CMB and late Universe e.g. galaxy clustering and lensing, probes of the evolution of matter perturbations \cite[see e.g.][]{2016JCAP...01..007B,2018JCAP...01..022B,2021PhRvD.103b3504D,2021arXiv210709664X}.

The LiMR can either be of thermal, or nonthermal origin. One plausible class of candidates for LiMRs are light \rm{eV} mass sterile neutrinos related to Short Base Line (SBL) anomalies\footnote{The new results of the MicroBooNE experiment \cite{MicroBooNE:2021jwr, Arguelles:2021meu} disfavour the sterile neutrino interpretation of the MiniBooNE anomaly as an electron neutrino appearance from a muon neutrino beam however a recent analysis shows a $2.2 \sigma$ preference for a sterile neutrino mass in the \rm{eV} scale if the MicroBooNE data are interpreted in terms of electron neutrino disappearance \cite{Denton:2021czb}.}. CMB and Large Scale Structure (LSS) observations strongly constrain the simplest scenario where the new light sterile neutrino component is a non-interacting and free-streaming species. In
such a minimal scenario, it is therefore very unlikely to find a  resolution to SBL anomalies at the same time being consistent with CMB and LSS data. That is why several models beyond the simple non-interacting thermal LiMR have been proposed - either with hidden interaction \cite{Archidiacono:2020yey} or with non-thermal or partially thermal distribution function \cite{Archidiacono:2013xxa}.

 At the same time, recent studies have found a persistent $2\sigma$-$3\sigma$ discrepancy between the predictions of late time clustering from the best-fit model to the \textit{Planck} CMB primaries \citep{2020A&A...641A...6P}, and actual measurements \cite[e.g.][]{2019PASJ...71...43H,2020JCAP...05..005D,2020Ivanov,2021A&A...649A..88T,2021arXiv210513549D,2021JCAP...12..028K}. This is usually framed in terms of differences in the value of $\sigma_8$ - the amplitude of matter fluctuations on $8\hmpc$ scale from linear perturbation theory - or in the value of  $S_8=\sigma_8(\Omega_m/0.3)^{0.5}$. While the \textit{Planck} and low redshift results are by no means statistically irreconcilable, the fact that all the low redshift analyses systematically find a lower value of $\sigma_8$ makes it an interesting problem. Since LiMR models naturally damp power on small scales, they offer a physical path toward reconciling the $\sigma_8$ discrepancy. Therefore, a thorough study of LiMR phenomenology in structure formation is warranted, both from the particle physics, as well as the cosmological, perspectives.

While the effects of LiMRs on structure formation, and possible constraints on LiMR models from survey data, have been quite extensively explored for large (linear) scales, their precise effects on nonlinear scales ($\lesssim 10 \hmpc$), especially for nonthermal LiMRs, remain uncharted --- see \citet{2017JCAP...10..015B} for small scale effects of thermal LiMRs\footnote{For mixed dark matter cosmologies with somewhat heavier relics, see, e.g., \cite{2009JCAP...05..012B, 2012JCAP...10..047A,2013JCAP...03..014A,2016PhRvD..94b3522K,2021JCAP...12..044P}.}.  Given the strong theoretical motivations for LiMRs and the upcoming observation that will be probing these scales,  these effects deserve a detailed study. The $\sigma_8$ discrepancy also provides a direct phenomenological motivation for characterizing the nonlinear effects of LiMR models.  In light of the discrepancy, it is important to understand the degeneracy between the effects of LiMRs and an overall rescaling, or a change in the tilt, of the initial power spectrum within the $\Lambda$CDM paradigm. Both produce a change in the linear $\sigma_8$, and therefore, effects on smaller scales could be key in distinguishing the models. Furthermore, the nonlinear scales can also, in principle, distinguish between various models of LiMR with different velocity distributions. In linear theory, the evolution of perturbations is determined by the effective velocity dispersion, i.e. the second moment of the velocity distribution, and not by the overall distribution. This implies that different LiMR models can produce the same effects on large scales, as long as their velocity dispersions match, even if their actual distributions are significantly different. On nonlinear scales, in contrast, clustering of the LiMR component can be sensitive to the details of the velocity distribution, offering a way to differentiate between models.

Taking these important points forward, in this paper, we use cosmological $N$-body simulations to follow the nonlinear evolution of LiMR cosmologies, directly modeling both the LiMR component and the CDM+baryon component,
and study the effects on certain cosmological observables relevant for current and future photometric surveys, such as the Dark Energy Survey\footnote{https://www.darkenergysurvey.org/} and the Legacy Survey of Space and Time (LSST) 
at the Vera Rubin Observatory (VRO)\footnote{https://lsst.org/}. We will consider LiMRs with explicitly nonthermal velocity distributions, as well as those which can be mapped onto thermal distributions.

   Given a large number of LiMR models, to focus our analysis for this ``First study" we use a particular model in the main text. This is the one put forward in \citet{Bhattacharya:2020zap, Das:2021pof}. This model is particularly relevant given its implications for the $\sigma_8$ tension \cite{Das:2021pof}. We compare and contrast the predictions from the LiMR model with the \textit{Planck} best-fit $\Lambda$CDM model. We also compare the nonlinear predictions from the LiMR model with that of an $\Lambda$CDM universe with an overall rescaling of the linear power spectrum such that its $\sigma_8$ matches that of the LiMR model. The methods developed easily generalize to LiMRs with any velocity distribution as long
   as they have negligible interactions. In Appendix \ref{sec:velocity_distribution}, we show how our results and findings generalize to other LiMR models explored in the literature, e.g., LiMRs following a Dodelson-Widrow distribution \citep{PhysRevLett.72.17}, and under what conditions their predictions differ.

The layout of the paper is as follows: in Sec. \ref{sec:style} we recap the main features of  LiMR models and briefly discuss their linear signatures. In Sec. \ref{sec:sims}, we discuss the implementation of such a model in an $N$-body framework. In Sec. \ref{sec:results}, we present the results from the $N$-body simulations and point out the differences between the LiMR model and the standard $\Lambda$CDM model. Finally, in Sec. \ref{sec:discussion}, we summarize the main results and discuss various interesting aspects of the study.

\section{Sterile LiMRs and their Impact on Linear Cosmology}

\label{sec:style}

  LiMRs with negligible interactions are completely characterized by their distribution functions and their mass. In the linear theory, the physics is sensitive only to the first two moments of their distribution functions  \cite[see e.g.][]{Acero:2008rh} and all observables are determined by two associated parameters. The first of these is the effective mass parameter $(m_{\rm eff})$, a measure of the contribution of the sterile particles to the present-day energy density:
\begin{equation}
\frac{m_{\rm eff}}{94.05 {\rm eV}} \equiv \left[
{{m_{\rm sp}} \int \!\! dp \,\, p^2 f(p)} 
\right] \times \left[\frac{h^2}{\rho_c^0}\right],
\end{equation}  
 where $m_{\rm sp}$  is the physical mass of the sterile particle and $\rho_c^0$ the total energy density of the universe today. The second parameter is
$\Delta N_{\rm eff}$, which measures the effective number of neutrino-like relativistic degrees of freedom at the time of neutrino decoupling:
\begin{equation}
\Delta N_{\rm eff} \equiv \left[
{ \int \!\! dp \,\, p^3 f(p)} 
\right] /
\left[{\frac{7}{8} \frac{\pi^2}{15}
{T_\nu}^4} \right]  
\end{equation}
with $T_\nu \equiv (4/11)^{1/3} T_{\rm cmb}$, the present-day temperature of the cosmic neutrino background.  The fact that the linear cosmology  is determined by
just two parameters implies that the CMB and linear matter power spectra are not powerful probes of specific models of LiMRs. 
As discussed in the introduction,   the goal of this paper is to analyze non-linear signatures and thereby address this shortcoming.

  Let us briefly review the  LiMRs distribution functions that will be relevant for our discussion.

\noindent \underline{\it{The Dodelson-Widrow distribution }}: The Dodelson-Widrow momentum distribution takes the form \cite{PhysRevLett.72.17}:
$$
f(p)=\frac{\chi}{\rm{exp}(p/T_{\nu})+1} \ , 
$$
where $\chi$ is a model parameter. The distribution function can be shown to be equivalent to a thermal distribution function at a temperature different
from $T_{\nu}$ by a redefinition of the parameters. This distribution is widely used in the context of sterile neutrino relics.

\noindent \underline{\it{ A Gaussian  distribution }}: Non-equilibrium decays can lead to distributions functions which are approximately Gaussian \cite[see e.g.][]{Cuoco:2005qr}:
 \begin{align}
 f(\vec{p}) =& N { T^{3}_{\nu} \over |\vec{p}|^2 } { \rm{exp} }\left(  -  \left( { ({|\vec{p}|   - p_0})^2 \over 2 \sigma^2  }\right)  \right); \nonumber \\
   N =&  \left( 4 \pi \sqrt{ {2  \pi \sigma^2 }} \right)^{-1}.\, \nonumber
\end{align}
 where $p_0$, $\sigma$ are model parameters.

\noindent \underline{\it{A Non-thermal distribution from decays:}} LiMRs with a non-thermal distribution can also be produced from the decay of a heavy scalar  \citep{1212.4160, 1304.1804, 1908.10369, Bhattacharya:2020zap}. 
In this case, the explicit form of the distribution function is a bit complicated. Here, we provide a qualitative description of the production process and refer the reader to the above references for the precise form
of the distribution function. At early times, the energy density of the universe is dominated by cold particles of a species $\varphi$. $\varphi$ decays to the 
Standard model sector and a light sterile particle (the LiMR). The Standard model sector thermalizes but the sterile particles do not. 
The production rate of the LiMRs is determined by the decay rate of the $\varphi$ particles and the branching ratio to the LiMR channel. The late-time momentum distribution of the LiMRs is determined
by redshifting them from the time of production. The distribution function is characterized by the mass of the heavy particle $(m_{\varphi})$, its decay rate ($\tau$), and the branching ratio for
decay to the LiMRs ($B_{\rm sp}$).

\begin{figure}[htp!]
\centering
\includegraphics[scale=0.345]{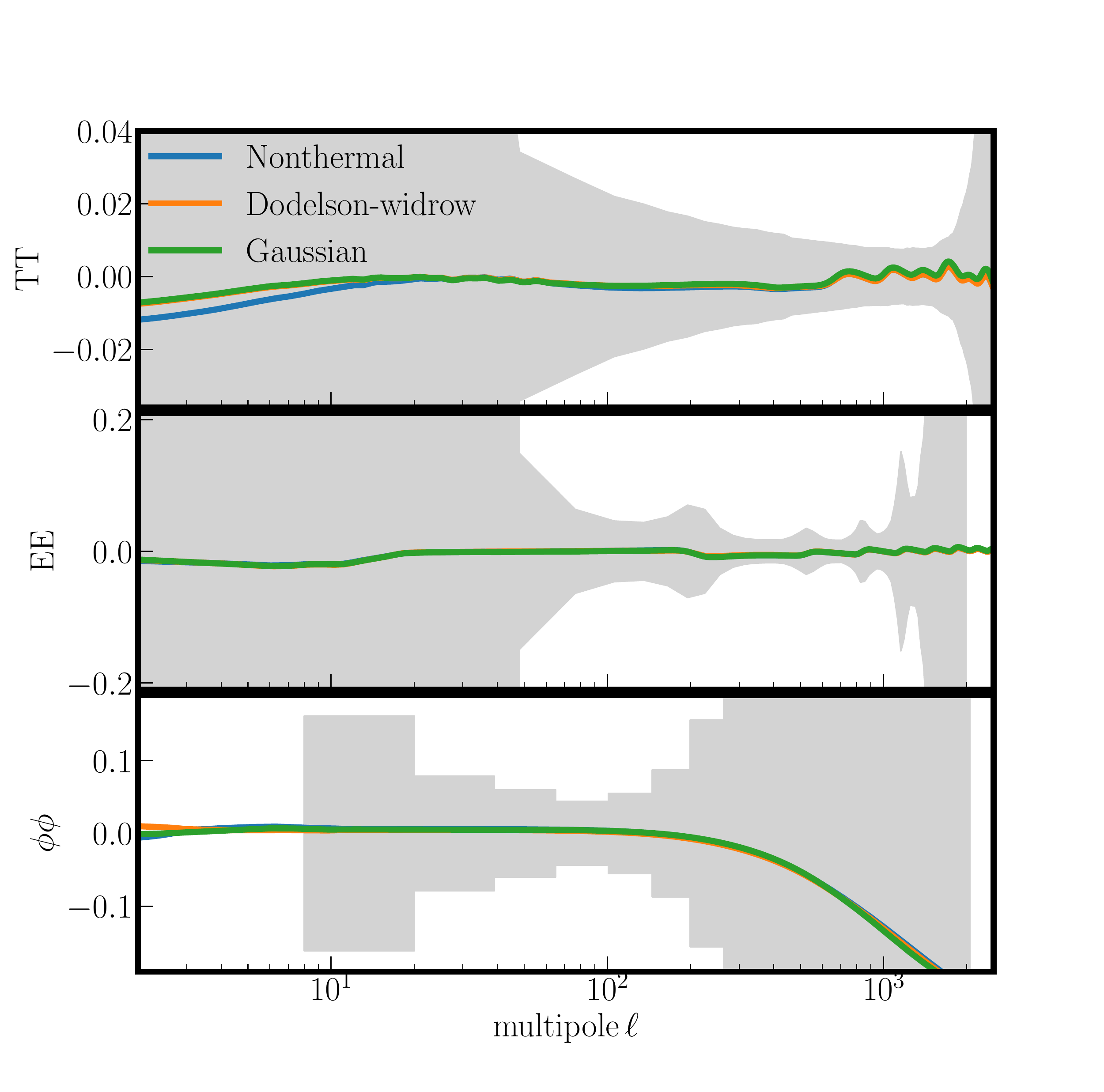}
\caption{The Residuals of CMB $C_l^{TT}$, $C_l^{EE}$, and $C_l^{\phi\phi}$ power spectra for all three models (see legend) to $\Lambda$CDM model using best-fit values of \planck data with a late-time $S_8$ prior. The shaded region is the Planck 2018 1$\sigma$ uncertainties. $\Delta N_{\rm eff}=0.034$ and  $m_{\rm eff}=0.90$eV in all three cases. All three models are indistinguishable from the baseline $\Lambda$CDM model given the \planck error bars.} 
\label{fig:nonthermal_cl} 
\end{figure}

\begin{figure}[htp!]
\centering
\includegraphics[scale=0.49]{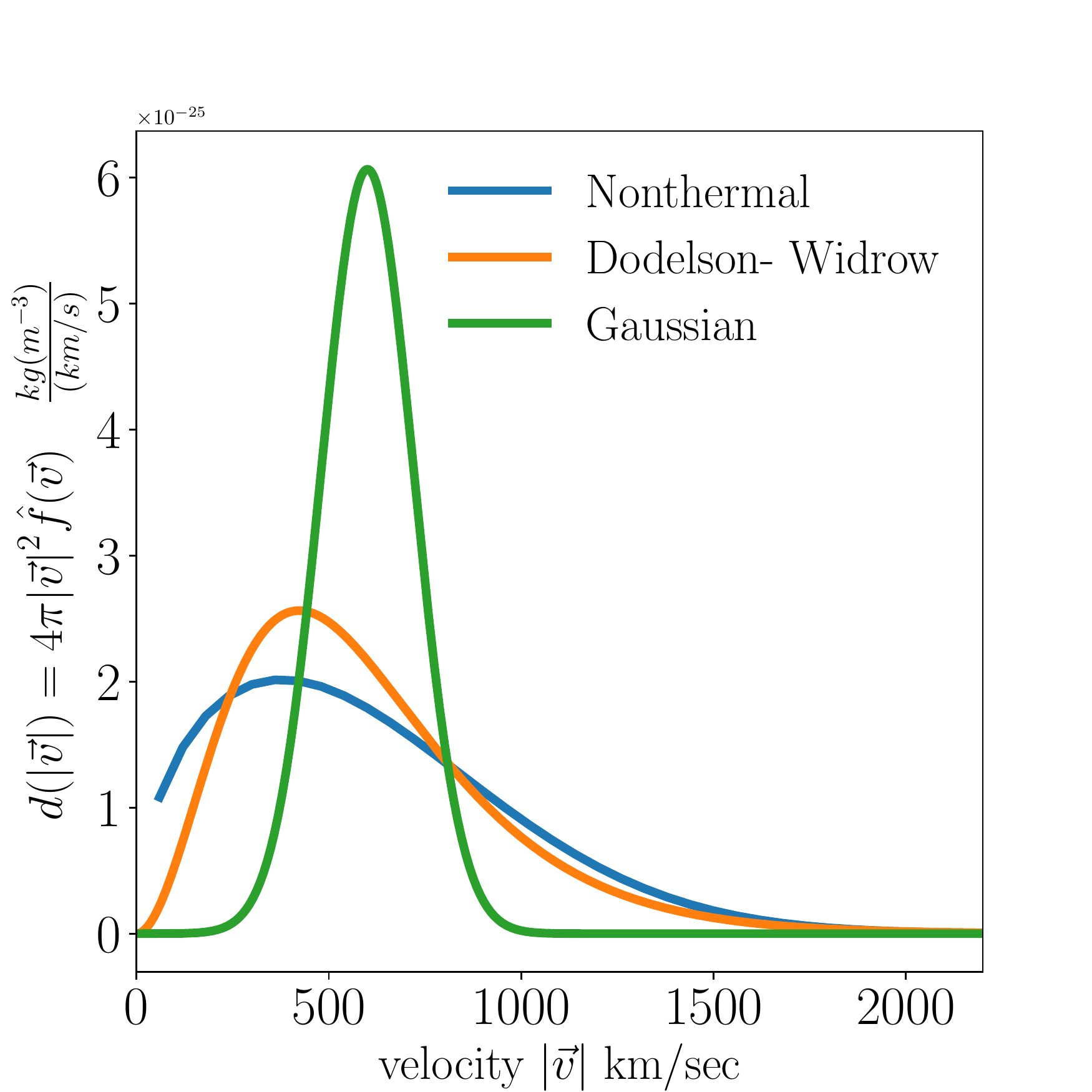}
\caption{The velocity distribution at $z=99$ of the Dodelson Widrow model, the Gaussian distribution and the  Nonthermal distribution from  decays. $\Delta N_{\rm eff}=0.034$ and  $m_{\rm eff}=0.90$eV in all three cases.} 
\label{fig:psd} 
\end{figure}

In \cite{Das:2021pof}, a  comprehensive Markov Chain Monte Carlo (MCMC) analysis for sterile LiMRs was carried out against up-to-date data from \planck \citep{2020A&A...641A...6P}, BOSS (BAO and $f\sigma_8$) \citep{Beutler:2011hx,Ross:2014qpa,Alam:2016hwk} and Pantheon data \citep{Pan-STARRS1:2017jku}, with and without the inclusion of a prior on the value of $S_8$ as measured with the KiDS/Viking+BOSS+2dFLens data \citep{Heymans:2020gsg}.  It was found that the tension between \planck and $S_8$ measurements can be alleviated by the presence of the nonthermal LiMR component.  The best fit values to \planck data  with a $S_8$ prior for the effective parameters were found to be 
$\Delta N_{\rm eff}=0.034$ and $m_{\rm eff}=0.90$eV. In terms of parameters relevant for the late-time Universe, these bounds translate to $\Omega_{\rm LiMR} \approx 0.021$ at $z=0$. These effective bounds on $m_{\rm eff}$ and $\Delta N_{\rm eff}$ can be translated to the parameters of the models discussed above\footnote{For the
Gaussian distribution and the nonthermal distribution from decays demanding particular values of $\Delta N_{\rm eff}$ and $m_{\rm eff}$ does not 
fix a point in their parameter spaces, but specifies  subspaces. For these models, we will choose particular points from the subspaces, leaving detailed
explorations of the full parameter spaces   for future work.}: for the Dodelson-Widrow Model, the best-fit values imply $\chi=0.034$ and $m_s=26.43\rm{eV}$. For the Gaussian distribution, they imply $p_0=0.01957T_{\nu}$, $\sigma={p_0}\big{/}{5}$ and $m_{\rm sp}=0.1644 \rm{eV}$. Finally, for nonthermal distribution from decays, $m_{\varphi} = 10^{-6} M_{\rm pl}$,  $\tau = 10^{8} \big{/} m_{\varphi}$,  $B_{\rm sp} = 0.0118$ and $m_{\rm sp} =38.62 \rm{eV} $. The matching of the effective parameters implies that the three models are indistinguishable at the linear level.
The residuals of the CMB $C_l^{TT}$, $C_l^{EE}$, and $C_l^{\phi\phi}$ for all three models to baseline $\Lambda$CDM model using best-fit values of \planck and $S_8$ data are shown in Fig.~\ref{fig:nonthermal_cl}.  All residual lie within Planck 2018 1$\sigma$ uncertainties and as expected, confirming that the models are statistically indistinguishable.

The key input for computing the effect of  LiMRS at the non-linear level is the normalized velocity distribution ($\hat{f}(\vec{v})$) at $z=99$ when the simulations are typically initialized, as described in more detail in the next section. This is easily computed from their momentum distribution functions and
is exhibited for the three models in  Fig. \ref{fig:psd}. Note that this is the background velocity distribution, and does not include the component sourced by the gravitational evolution of perturbation.

   To keep the discussion of the paper focused, we will work with the non-thermal distribution from decays discussed in \citep{Bhattacharya:2020zap, Das:2021pof} in the main text of the paper. The model parameter
will be taken to be described above, i.e, best fit to Planck with a $S_8$ prior. In Appendix~\ref{sec:velocity_distribution}, we will report the results for the other models and discuss the prospects of distinguishing between the models
by making use of non-linear signatures.

\section{Simulation setup}
\label{sec:sims}

\begin{figure*}[htp]
    \centering
    \includegraphics[scale=0.5]{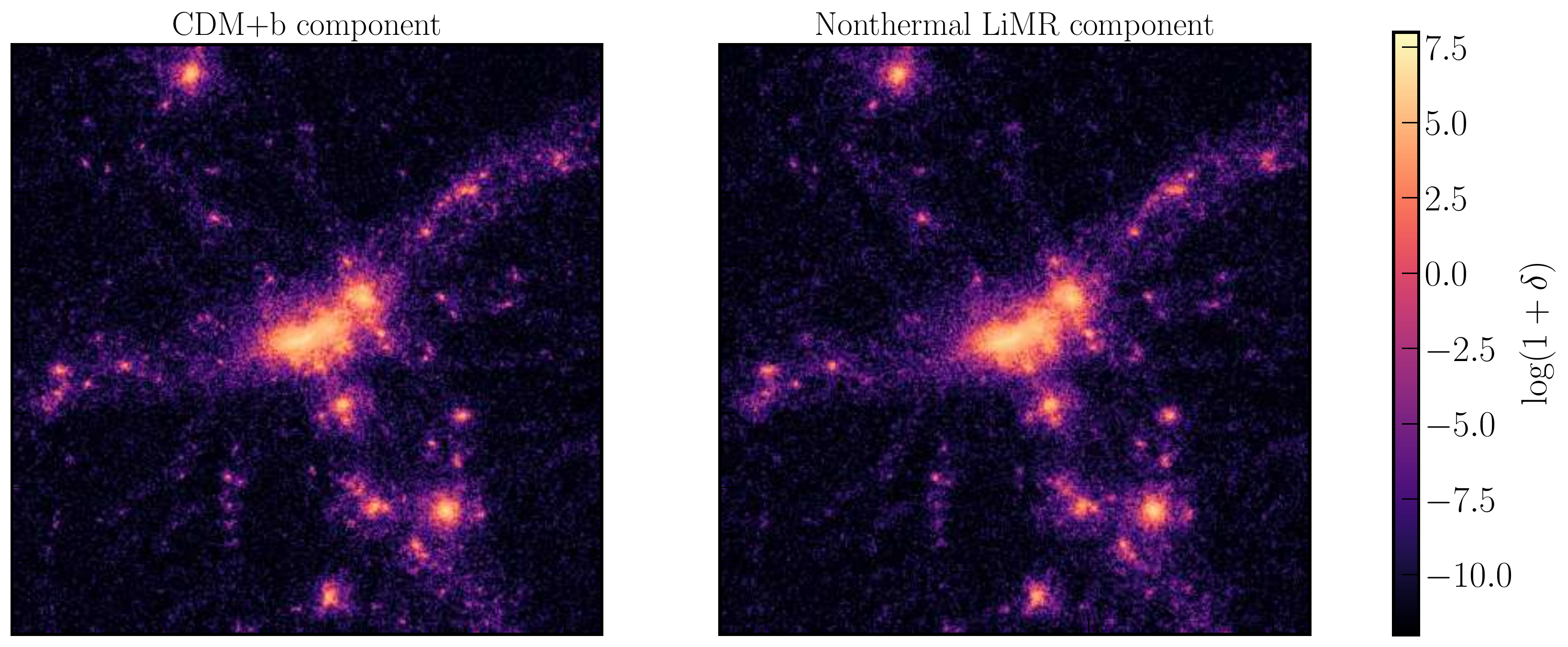}
    \caption{Projected density fields from a $(30\hmpc)^3$ volume of the simulation from the model with a LiMR component. The left panel shows the projected density field for the CDM+b component in the simulation. The right panel shows the projected density field of the LiMR component. While the cosmic web structure is clear in both components, the LiMR structures are more smeared out.}
    \label{fig:projection}
\end{figure*}

To model the evolution of the matter perturbations in the LiMR models into the nonlinear regime at low redshifts, and to contrast the results with those from the \textit{Planck} best-fit $\Lambda$-CDM model, we make use of $N$-body simulations. We use a suitably modified version of the publicly available cosmological $N$-body code \textsc{Gadget2} \citep{2005MNRAS.364.1105S}. In simulations of the LiMR model, we make use of two particle types --- one type representing the mass-weighted average of the CDM and baryon components, and the other representing the LiMR component. The particle masses are set by the background abundances of the different components. We note that the standard neutrinos are not actively modeled in the simulations used in this paper, they are only accounted for in the background evolution. We adopt such a simplifying approximation for this first study since the SM neutrino mass is the same between both models, and because the size of the LiMR effects are much larger than that of the SM neutrinos (typically $\sim 5\%$ for $M_\nu=0.06{\rm eV}$ as adopted here), as we demonstrate in the next section. 

To generate the initial conditions, we use the linear power spectrum and growth rate of various components at $z=99$ from the publicly available Boltzmann code \textsc{CLASS} \citep{2011JCAP...07..034B}. The modifications needed to correctly include the LiMRs, with different velocity distributions, within \textsc{CLASS} have been tested and discussed in \cite{Das:2021pof}. These are then converted into the initial positions and velocities of particles in the $N$-body simulation through the Zel'Dovich approximation \citep{1970A&A.....5...84Z} implemented in the \textsc{NGenIC} code\footnote{https://www.h-its.org/2014/11/05/ngenic-code/}. To account for the fact that the $N$-body evolution does not take into account the radiation component, while the \textsc{CLASS} outputs do  \cite[see e.g.][for a detailed discussion]{2017MNRAS.466.3244Z}, we use an overall rescaling in the $z=99$ power spectrum to ensure that the simulations give the correct linear growth rate on large scales at $z=0$.

For the CDM simulation particles, the initial positions and velocities are determined purely from the $z=99$ power spectrum and growth rate. However, for the simulation particles representing the LiMR component, the non-negligible velocity distribution, discussed in Sec. \ref{sec:style}, also needs to be taken into account. For SM neutrinos, there exists a large body of literature on the various methods for including the velocity distribution in cosmological $N$-body simulations, and their merits and drawbacks \cite[see e.g.][ and citations therein]{2008JCAP...08..020B,2010JCAP...06..015V,2013MNRAS.428.3375A,2016JCAP...11..015B,2018JCAP...09..028B,2021JCAP...01..016B,2021MNRAS.507.2614E}. In this paper, we adopt an approach similar to the one used in \cite{2010JCAP...06..015V} to assign velocities to the LiMR particles. The procedure is briefly summarized as follows: given a theoretical velocity distribution $f(v)$, we compute the Cumulative Distribution Function ${\rm CDF}(v)$. By definition, ${\rm CDF}(v)$ takes values between $0$ and $1$. For each LiMR particle in the simulation, we generate a random number, $x$, from an underlying uniform distribution over the interval $(0,1)$. The velocity $v^*$ for which ${\rm CDF}(v^*)=x$ is chosen as the magnitude of the velocity assigned to the particle. In addition, a random direction in the simulation box is chosen, and this velocity is added (as 3-vectors) to the gravitational peculiar velocity obtained from the linear $P(k)$ and growth factor. This procedure is repeated for every LiMR particle in the simulation\footnote{In principle, the velocity distribution of the LiMR can be a function of the local potential \cite[see e.g.][]{2018MNRAS.481.1486B}, but at high enough redshifts, when the gravitationally sourced peculiar velocities are small compared to the typical LiMR velocity, the approximation of a position-independent velocity distribution is justified.}.

With this setup, we run two simulations - one with the nonthermal LiMR component, and the other for the $\Lambda$CDM best fit over $(1\hgpc)^3$ volumes. We run a separate $\Lambda$CDM box for which the value of $\sigma_8$ at $z=0$ is matched to the nonthermal LiMR cosmology. For the $\Lambda$CDM simulations, we use $1024^3$ particles to represent the CDM and baryonic components. The mass of the simulation particles is $7.92\times 10^{10}M_\odot/h$. For the simulations with nonthermal LiMRs, we again use an additional $1024^3$ particles to represent this component. The mass of the ${\rm cdm+b}$ simulation particles is $7.51\times 10^{10}M_\odot/h$, and the LiMR simulation particles is $5.51\times 10^{10}M_\odot/h$ in this simulation.  The force softening scale for both simulations is set to $25\hkpc$. In Fig. \ref{fig:projection}, we show the projected density fields from the simulation with the LiMR component. In this simulation, there are two distinct fields - the CDM+baryon component shown on the left, and the LiMR component on the right. The initial velocity distribution of the LiMR component leads to ``fuzzier" structures, characterized by less prominent peaks and voids, compared to the CDM+baryon component. We use the \textsc{Rockstar} halo finder \citep{2013ApJ...762..109B} to identify the positions and masses of dark matter halos. It is worth noting here that \textsc{Rockstar} identifies halos using only the \textsc{Gadget} Type $1$ particles, i.e., those corresponding to the CDM+baryon component in the simulations. We use $M_{\rm vir}$ and $R_{\rm vir}$ reported by \textsc{Rockstar} to represent the mass and radius of the halos.

\begin{figure}[htp!]
\centering
\includegraphics[scale=0.45]{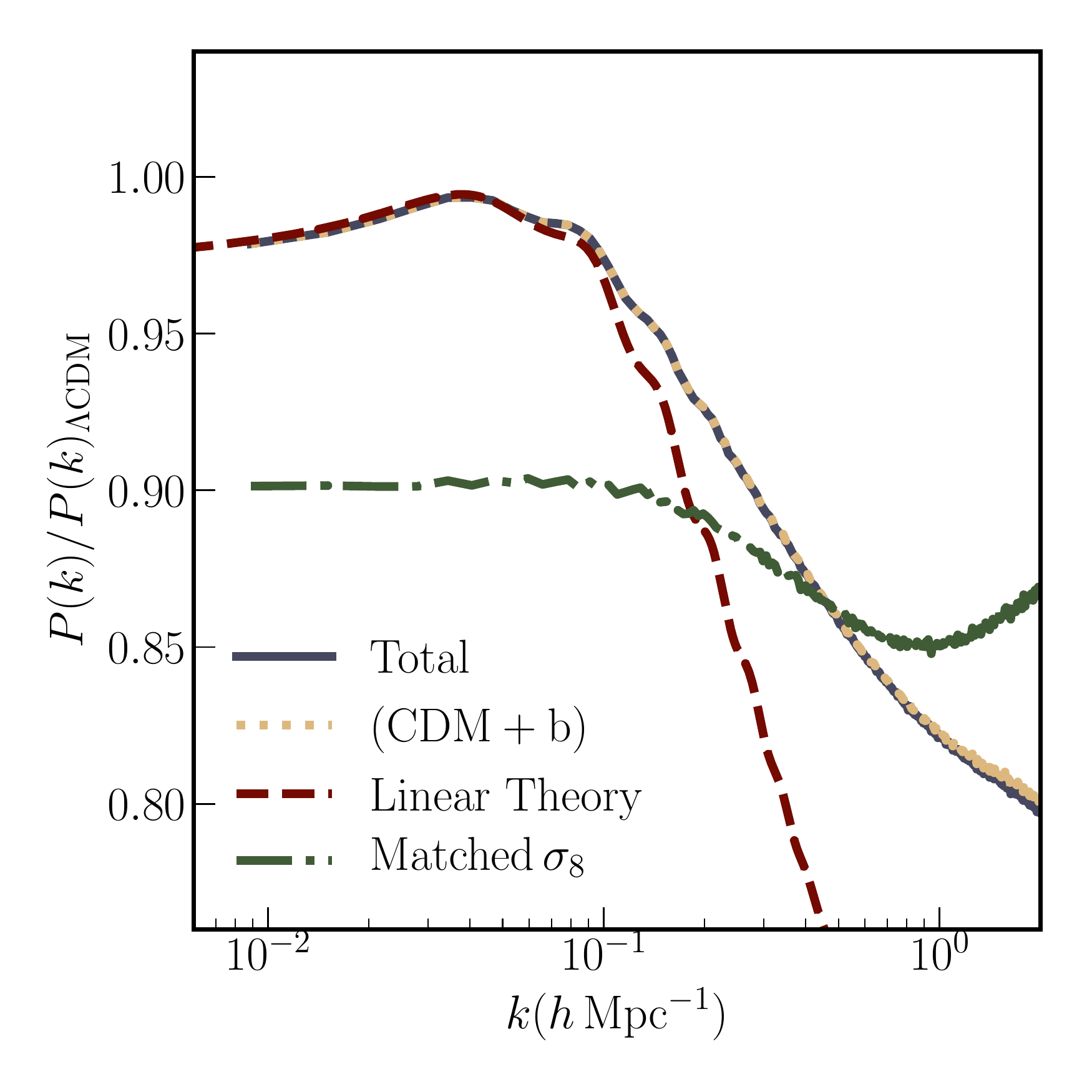}
\caption{Ratio of power spectra, with respect to the $\Lambda$CDM Planck best-fit model. The (solid) blue line represents the ratio for the total matter power spectrum from the nonthermal LiMR model. The (dashed) maroon line represents the same ratio, but predicted by linear theory (\textsc{CLASS}). The (dashed) yellow line represents the ratio for the CDM+baryon power spectrum from the nonthermal LiMR model. Finally, the (dot-dashed) green line represents the ratio for a $\Lambda$CDM model, but with $z=0$ $\sigma_8$ matched to the nonthermal LiMR model. While the nonlinear damping is smaller than the one predicted by linear theory for the nonthermal LiMR model, the shape is significantly different, from that produced by an overall amplitude rescaling within the $\Lambda$CDM model.} 
\label{fig:pk} 
\end{figure}

\section{RESULTS }
\label{sec:results}

In this section, we present the results from the $N$-body simulation, and discuss the differences between the best-fit $\Lambda$CDM model and the nonthermal LiMR model, focusing on three different ``observables". As mentioned previously, we focus specifically on the results of simulations with the nonthermal LiMR model \citep{Bhattacharya:2020zap} in this section. We refer the reader to Appendix \ref{sec:velocity_distribution} for a comparison with other LiMR models defined by their velocity distributions.

\subsection{Power Spectrum}
\label{sec:pk_damping}
First, we focus on the comparison of the nonlinear power spectrum at $z=0$ from the $\Lambda$CDM simulation and the nonthermal LiMR simulation. For the nonthermal LiMR cosmology, multiple different power spectra can be defined. The total matter power spectrum, $P_m(k)$, is defined as 
\begin{equation}
    P_m(k) = f_{cb}^2 P_{cb}(k) + 2f_{cb}f_{nt} P_{cb,nt} + f_{nt}^2 P_{nt} \, ,
\end{equation}
where the subscript $cb$ refers to the CDM+baryon component, and the subscript $nt$ refers to the nonthermal LiMR component, and
\begin{equation}
    f_{cb} = \frac{\Omega_{cb}}{\Omega_{cb}+\Omega_{nt}} \, \, ; \, \, f_{nt} = \frac{\Omega_{nt}}{\Omega_{cb}+\Omega_{nt}} \, .
\end{equation}
For this specific model, $f_{nt}=0.068$ and $f_{cb}= 0.932$. $P_{cb}(k)$ is the auto power spectrum of the CDM+baryon component, $P_{nt}$ is the auto power spectrum of the nonthermal component, and $P_{cb,nt}$ is the cross power spectrum of the two components. Note that observables such as weak and strong lensing, which depend on the total gravitational potential, is sensitive directly to $P_m$, whereas the clustering of galaxies, whose locations are determined by peaks of overdensities in the CDM+baryon component is sensitive directly to $P_{cb}$. The other two components, $P_{nt,cb}$ and $P_{nt}$ do not correspond directly to any cosmological observable. For a pure $\Lambda$CDM cosmology, $P_m$ and $P_{cb}$ are the same, but in massive neutrino cosmologies, for example, the difference can be important \citep{2014JCAP...03..011V,2016JCAP...11..015B,2020JCAP...06..032B}.
 
In Fig. \ref{fig:pk}, we plot the ratio of various $P(k)$ to the $P(k)$ obtained from the Planck best-fit $\Lambda$CDM simulation. The (solid) blue line represents the ratio for the total $P_m(k)$ from the nonthermal LiMR simulation, while the (dotted) yellow line represents the ratio for $P_{cb}$. For reference, the linear theory prediction for the ratio of $P_m(k)/P(k)_{\Lambda{\rm CDM}}$ is plotted with the (dashed) maroon line. The damping of both $P_m$ and $P_{cb}$ on small scales is clear, but the amplitude of the damping is significantly lower compared to the prediction from linear theory. This is consistent with the findings of \citet{2017JCAP...10..015B} in the context of thermal LiMRs. Even then, the resultant damping of $\sim 15\%$ at $k\sim 1 \ihmpc$ is much larger than the damping produced by standard neutrinos with $M_\nu=0.06$eV. Also in contrast to standard neutrinos, where the damping starts to become significant from $k\sim 10^{-2}\ihmpc$, the damping in this case starts at $k\sim 10^{-1}\ihmpc$. The difference in the damping scale, related to the free streaming scale of the LiMR, arises from the difference in the velocity distribution of the nonthermal LiMRs from that of standard neutrinos. The (dot-dashed) green line in Fig. \ref{fig:pk} represents the ratio for the $\Lambda$CDM cosmology with (linear) $\sigma_8$ matched to the nonthermal LiMR cosmology. As expected, the ratio is scale independent on large scales, but importantly, has a different shape on small scales than that from the nonthermal LiMR case. This is again qualitatively different from the behavior seen in standard massive neutrino cosmologies, where the small scale damping of $P(k)$ is roughly degenerate between $M_\nu$ and $\sigma_8$ \cite[see e.g.][for more detailed discussion]{2021arXiv210804215B}. 

In a photometric survey, which is sensitive to both $P_{cb}$, through galaxy clustering, and to $P_m$, through cosmic shear or galaxy-galaxy lensing, the small scale damping signal should be measurable given sufficient sensitivity on those scales. In addition, the fact that this damping is not degenerate with a change of $\sigma_8$ makes it easier to rule out or detect this model using a combination of galaxy clustering and lensing. We leave a quantitative estimate of the constraints from these observables to a later work.

\begin{figure}[htp!]
\centering
\includegraphics[scale=0.45]{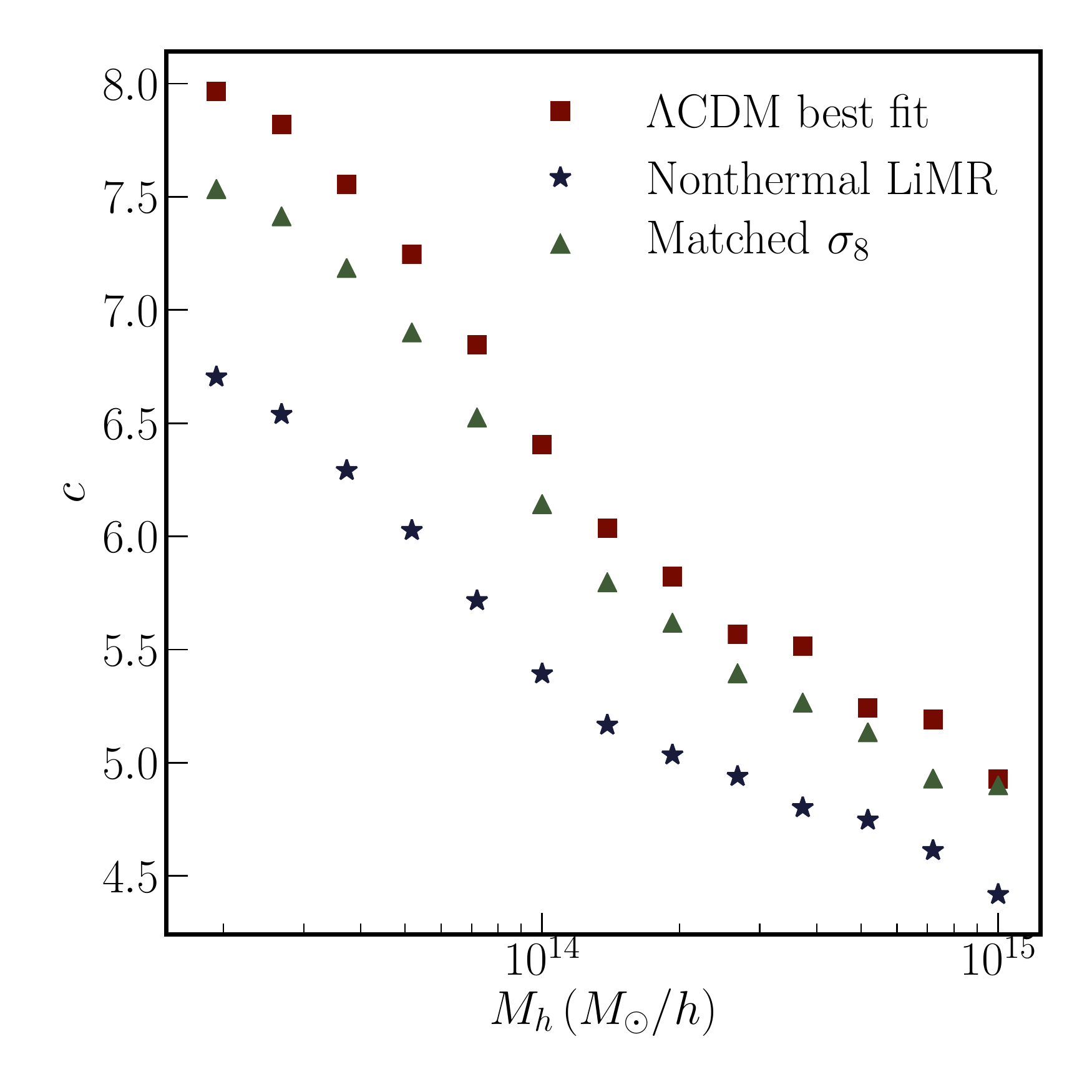}
\caption{Mean mass-concentration relation obtained for the three models. The maroon (square) points are for the \textit{Planck} best-fit $\Lambda$CDM model, the blue (star) points are for the nonthermal LiMR model, while the green (triangle) points are for the $\Lambda$CDM model with $\sigma_8$ matched to the nonthermal LiMR model. The nonthermal LiMR model has a marked effect, lowering the mean concentration at fixed mass, and a simple change of $\sigma_8$ within the $\Lambda$CDM model is unable to replicate this.} 
\label{fig:mass_conc} 
\end{figure}

\subsection{Mass-concentration relation}
\label{sec:mass_conc_relation}

The evolution history of dark matter halos over cosmic time is encoded in various properties of its density profile. The average concentration of halos as a function of the virial mass termed the mass-concentration relationship, is one such property \citep{1996ApJ...462..563N,2001MNRAS.321..559B}. The concentration ($c$) is defined in terms of the ratio between the virial radius of the halo ($r_{\rm vir}$), and the scale radius of the halo ($r_s$), i.e. $c = r_{\rm vir}/r_s$. The scale radius itself is defined as the radial scale at which the logarithmic derivative of the density profile of the halo, ${\rm d \, }\log \rho/{\rm d \,} \log r $ is $-2$. Considerable observational efforts have been made to determine the mass-concentration relation in the Universe, using a variety of probes. Gravitational lensing, both in the strong lensing and weak lensing regime, is one such probe \cite[see e.g.,][]{2007MNRAS.379..190C,2008JCAP...08..006M,2014ApJ...784L..25C,2014ApJ...795..163U,2016ApJ...821..116U,2015ApJ...814..120D,2015ApJ...806....4M,2016A&A...586A..43V}, and will continue to yield more precise measurements of the mass-concentration relation with wider area photometric surveys.

In Warm Dark Matter (WDM) cosmologies, where the power spectrum is damped on small scales due to the thermal motion of the dominant DM component, the mean relationship between the mass and concentration of dark matter halos is found to be different from that seen in a Cold Dark Matter model \citep{2012MNRAS.424..684S,2013MNRAS.428..882M,2016MNRAS.460.1214L}. Heuristically, this happens because the damping delays the onset of nonlinear evolution on small scales, and low mass halos form later than in a CDM scenario. Since the concentration of halos is correlated with the density of the background universe at the time of formation, delayed formation times imply a lower background density, and hence, concentrations. Therefore, unlike in CDM scenarios, where the mean concentration continues to rise as one moves to lower mass, WDM models produce a non-monotonic relationship, with the mean concentration decreasing to both sides of a particular halo mass scale (set by the WDM particle mass). The nonthermal LiMR model considered here also produces a damping of power on small scales, but due to a completely different mechanism from WDM models --- in this case, the presence of a subdominant but the non-negligible component that clusters less than the CDM component. Therefore, we investigate if it produces an analogous effect on the mass-concentration relation. In WDM models with realistic masses, the damping starts on scales of $k > 1 \hmpc$, and consequently, has little effect on the mass-concentration relation for cluster and group sized halos ($\gtrsim 10^{13} M_\odot/h$). In the nonthermal LiMR model, the damping, relative to the CDM model, starts at much larger scales, as shown in Fig. \ref{fig:pk}, and can potentially affect larger mass halos.

To obtain the mean relationship between mass and concentration, we use the halo mass reported by \textsc{Rockstar} and bin the halos into 20 bins (logarithmic) between $10^{13} M_\odot/h$ and $10^{15}M_\odot/h$. For each halo, we also compute the concentration ($c$) from the values of $R_{vir}$ and scale radius $R_s$ reported by \textsc{Rockstar}, using $c = R_{vir}/R_s$. We then compute the mean concentration of halos in a given mass bin. Note that for the lowest mass bin, the number of particles in each halo is $\sim 200$, and therefore the \textsc{Rockstar} estimates for $R_s$ may not be entirely converged. However, this should affect all the simulations in roughly the same way. Results from the various simulations are plotted in Fig. \ref{fig:mass_conc}. The maroon (square) data points represent the mass-concentration relationship for the \textit{Planck} best-fit $\Lambda$CDM model, the blue (starred) data points for the nonthermal LiMR model, and the green (triangular) data points for the $\Lambda$CDM model with $\sigma_8$ matched to the nonthermal LiMR model. The nonthermal LiMR model, on the scales considered here, does not seem to produce a non-monotonic mass-concentration relation as seen in WDM, but instead, an overall reduction in the mean concentration over the entire range of mass scales. This includes objects at the very high end of the mass scale, i.e. massive galaxy clusters. It is interesting to note that changing $\sigma_8$ within the $\Lambda$CDM paradigm, i.e. the green data points, produce a much smaller effect. Since the mass-concentration relation depends on the full evolutionary history, merely matching $\sigma_8$ at $z=0$ is not sufficient to capture the effects of the nonthermal LiMRs. While the mean relation remains monotonic over the range of halo mass scales that are well resolved in these simulations, it will be interesting to study the behavior at lower mass scales, and possible implications for smaller systems, including for the satellite population of a system like the Milky Way. This will require higher resolution simulations, and we leave this to future work.

\subsection{Weak lensing around massive clusters}
\label{sec:cluster_lensing}

\begin{figure}[htp!]
\centering
\includegraphics[scale=0.48]{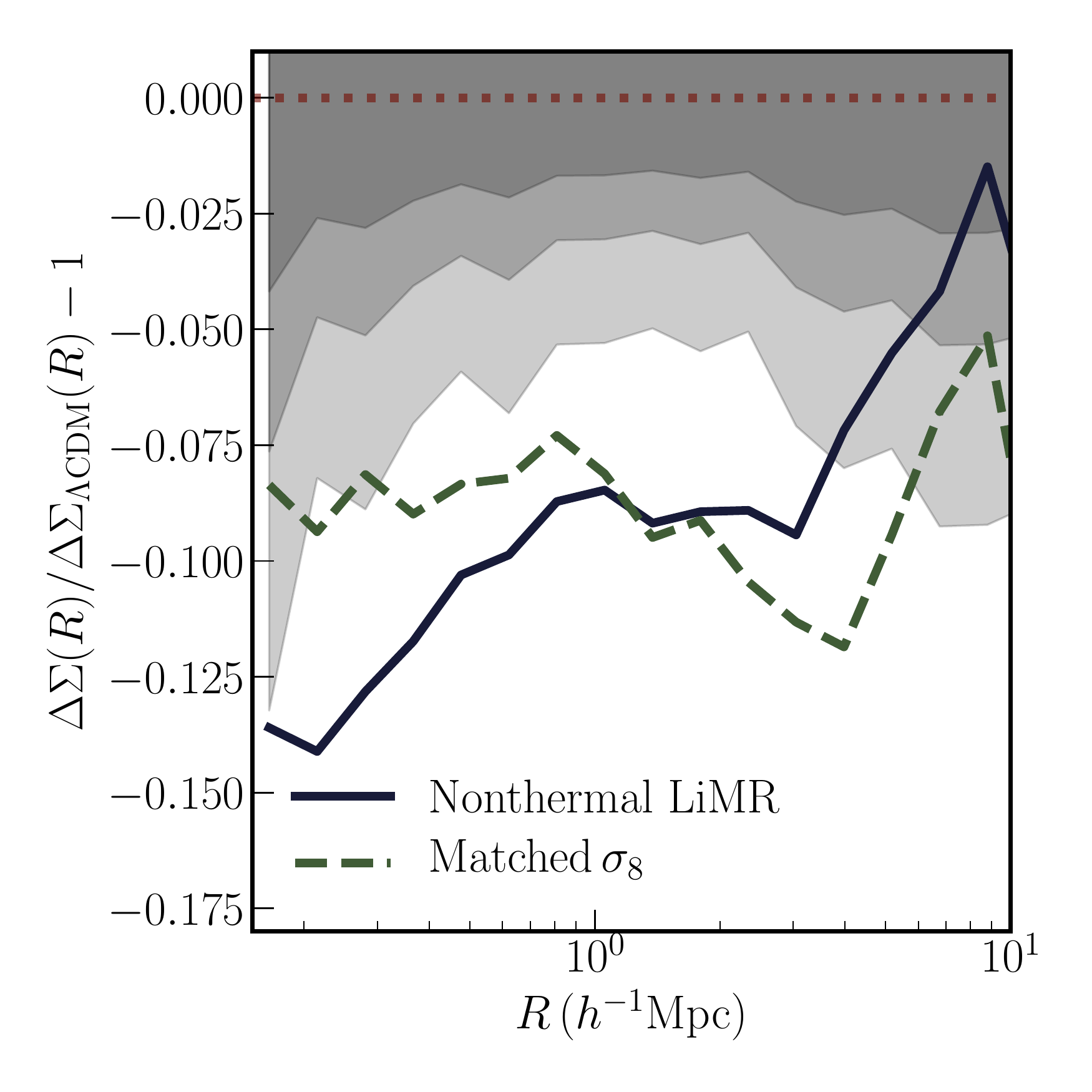}
\caption{Residual of $\Delta \Sigma$, with respect to the $\Lambda$CDM Planck best-fit. The maroon (dotted) line represents $0$ residual. The blue (solid) line represents the residual for the nonthermal LiMR model. The green (dashed) line represents the residual for a $\Lambda$CDM model with $\sigma_8$ matched to the nonthermal LiMR model. The lightest shaded region corresponds to error estimates from the measurements in \citet{2018ApJ...864...83C} in DES Y1 data, rescaled to the measurement of the signal in the \textit{Planck} best fit $\Lambda$CDM model (see text for more details). The progressively darker shaded regions represent projections for error bars expected in DES Y3 and LSST (VRO), based on the increase in the fraction of the sky covered by those surveys compared to DES Y1.} 
\label{fig:delta_sigma} 
\end{figure}

Finally, we consider the stacked weak lensing measurements around cluster-mass halos, typically for halo masses $\gtrsim 10^{14}M_\odot/h$. In recent years, such measurements have become increasingly accurate and precise \cite[see e.g.][]{2016PhRvL.116d1301M,2018ApJ...864...83C,2021MNRAS.507.5758S}, and is a sensitive probe of clustering on scales $0.1 \hmpc \lesssim r \lesssim 10 \hmpc$. Since these small scales are affected most by the presence of the nonthermal component, cluster lensing can be an important tool to constrain or rule out such models.

The weak lensing signal measured in the data is related directly to the excess surface mass density
\begin{equation}
    \Delta \Sigma (R) = \Sigma(<R) - \Sigma(R) \,  ,
    \label{eq:delta_sigma}
\end{equation} 
where $R$ is the projected distance from the cluster center. In order to compute this quantity in the simulations, we first identify one of the directions in the simulation volume as the line-of-sight (LoS) direction. For each halo, we extract all particles in a cylinder with height $100\hmpc$ and axis along the LoS direction, centered at the halo center as identified by \textsc{Rockstar}. For a given $R$, the first term on the RHS of Eq. \ref{eq:delta_sigma} is computed by counting up the mass enclosed in the cylinder out to radius $R$ and dividing by the area enclosed. The second term on the RHS of Eq. \ref{eq:delta_sigma} is calculated by considering the mass in  this circular annulus around $R$. We compute $\Delta \Sigma(R)$ in logarithmically spaced bins between $0.15\hmpc$ and $15\hmpc$. We repeat the measurement around the center of the $20000$ most massive halos in each simulation box and compute the average (stack) of those measurements. One caveat to keep in mind here is that the mass of the halo is taken from the \textsc{Rockstar} catalog, which only uses Type 1 particles (representing the CDM and baryonic components) in its halo identification and mass measurements.

In Fig. \ref{fig:delta_sigma}, we plot the residual of the measurement of $\Delta \Sigma(R)$ from the measurement in the \textit{Planck} best-fit $\Lambda$CDM simulation. The (dotted) maroon line represents no deviation from the fiducial model. The (solid) blue line represents the residual for the nonthermal LiMR model, while the (dashed) green line represents the residual for the $\Lambda$CDM simulation with linear $\sigma_8$ matched to the nonthermal LiMR model. The lightest gray shaded region represents the relative $1\sigma$ error regions derived from measurements on a sample of halos from \citet{2018ApJ...864...83C} using the Dark Energy Survey (DES) Y1 data. The measurements were made on a sample of \textsc{redMaPPer} clusters \citep{2016ApJS..224....1R} with a richness cut of $20< \lambda <100$, and a redshift range of $0.2<z<0.55$. To translate the data error bars into meaningful error estimates for the simulation sample we consider in this paper, we rescale the data covariance matrix by the square of the ratio of the mean signal in the data and that obtained from the fiducial $\Lambda$CDM simulation. Note that this retains the structure of covariances between different radial bins, i.e. those encoded in the off-diagonal terms. It is also worth noting that the simulation measurements are at $z=0$, while the data covers a redshift range of $0.2<z<0.55$. Therefore, we are making the simplifying assumption that the error bars have a weak redshift dependence. The progressively darker gray regions represent the same errors rescaled to take into account the greater sky coverage expected in DES Y3 and LSST (VRO) data. Note, however, that the rescaled error bars do not take into account the deeper coverage, and hence the increase in the number density of background lensed galaxies. These shaded regions, therefore, serve as a rough guide for the degree to which the nonthermal LiMR model can be distinguished from the best-fit $\Lambda$CDM model in various surveys.

We find that the nonthermal LiMR model produces significantly different predictions, in terms of the shaded gray regions, for the stacked weak lensing signal around the $20000$ most massive halos in the simulations. In terms of the DES Y1 data covariance matrix, rescaled to the simulation mean as discussed above, the $\Delta \chi^2$ between the \textit{Planck} best fit $\Lambda$CDM model and the nonthermal LiMR model is $31.32$ for $16$ degrees of freedom. In terms of the projected DES Y3 error bars, the $\Delta \chi^2$ value is $93.64$, and for the projected LSST (VRO) error bars, it is $313.21$. For the last, the implied $p$-value is $<1e-5$, suggesting that the nonthermal LiMR model and the fiducial $\Lambda$CDM model can be distinguished at a very high level of statistical significance. While this is a rough calculation including multiple approximations, we have attempted to be as conservative as possible in terms of the expected signal-to-noise improvements in DES Y3 and LSST.

It should be noted that the difference in the lensing signal exists both in the $1$-halo virialized regime, i.e. $\lesssim 1 \hmpc$, but also extend out to the largest scales we have measured, i.e $10 \hmpc$ in the infall regime of these objects. This is in contrast to another extension of the $\Lambda$CDM model --- the Self-Interacting Dark Matter (SIDM) model, where the signal is affected only within the virial radius due to (elastic) scattering between dark matter particles \citep{2020JCAP...02..024B}. It is also worth noting that the effects of changing $\sigma_8$ are not degenerate with the presence of the nonthermal LiMR component, as seen by the difference between the blue and yellow curves. The difference between the two models is especially pronounced on small scales, $\lesssim 1 \hmpc$. This difference on small scales is especially relevant, since a number of studies pointing to a $\sigma_8$, or $S_8=\sigma_8(\Omega_m/0.3)^{0.5}$, tension between \textit{Planck} primary anisotropies and low redshift probes of clustering \cite[e.g.][]{2019PASJ...71...43H,2020JCAP...05..005D,2020Ivanov,2021A&A...649A..88T,2021arXiv210513549D,2021JCAP...12..028K}, typically do not use these smaller scales. Therefore, even if the nonthermal LiMR model is tuned to solve the $\sigma_8$ tension on intermediate, quasi-linear scales, the difference in the smaller scale, nonlinear regime predictions can be used to constrain or rule out the model. Once again, we refer the readers to Appendix \ref{sec:velocity_distribution}, where we demonstrate how the discussion above generalizes to other LiMR model classes.

\section{Discussion and summary }
\label{sec:discussion}

   In this paper, we have initiated a systematic study of nonlinear signatures of a mixed dark matter cosmology with CDM and a nonthermal LiMR component, focusing on those effects and differences from standard $\Lambda$CDM universes that have implications for observables at various current and future photometric survey. We use cosmological $N$-body simulations which actively model both the cold and LiMR components simultaneously. In the main text, the prototypical
model used for the first study is that of  \citet{Bhattacharya:2020zap, Das:2021pof}; preliminary studies involving other distribution functions were carried out in the appendix. Let us summarise our results:

\begin{itemize}

\item Firstly, we have explored the effects on the power spectrum, and have found a scale-dependent damping on small scales. The amplitude of damping is significantly different from that predicted by linear theory, validating the need for full $N$-body simulations to calibrate the effects of this model on small scales. Crucially, the shape of the damping is also different from that obtained by a simple rescaling of the initial power spectrum within the $\Lambda$CDM model. This implies that signatures on small scales can be used to discriminate between the two models, possibly breaking degeneracies that exist on larger scales.

\item Next, we have examined the effects on the mass-concentration relationship of dark matter halos, and find that the nonthermal LiMR model produces an overall reduction in the mean concentration as a function of halo mass over the mass range $10^{13}M_\odot/h < M_h < 10^{15}M_\odot/h$, but remains roughly monotonic. Once again, the level of the reduction cannot be reproduced by simply rescaling the $\Lambda$CDM power spectrum to match the $\sigma_8$ of the nonthermal LiMR model.  We have examined the signatures of the nonthermal LiMR model on the weak lensing measurements around the most massive galaxy clusters. We have shown, using certain simplifying assumptions, that the expected level of signal-to-noise in these types of measurement expected in DES Y3, and especially LSST (VRO), should be sufficient to discriminate between the nonthermal LiMR model and the \textit{Planck} best-fit $\Lambda$CDM model at a high level of statistical significance. To our knowledge, this is the first simulation work to study the effects of a class of LiMRs on cluster lensing, and demonstrate that it is possible to distinguish those classes of models from a standard $\Lambda$CDM universe using such measurements.

\item Finally, in the appendix we studied non-linear signatures of LiMR models which are completely indistinguishable at the linear level. Interestingly, we have found differences at the non-linear level. These can be considered as 
probes of the LiMR velocity distribution functions. And optimistically, since the velocity distributions of LiMRs are tied to their production mechanism, these can be a window
into early universe dynamics.

\end{itemize}

LiMR models can, therefore, be best tested and constrained by jointly analyzing data from CMB and late-time structure formation on both large \textit{and} small scales.

We now discuss certain interesting aspects of our results, and avenues for future work. One aspect, that we have highlighted throughout Sec. \ref{sec:results} is that the effects of the LiMR model lie somewhere between those seen for massive neutrinos and those in WDM cosmologies, without being completely degenerate with either. The scale at which the power spectrum starts damping, w.r.t to the $\Lambda$CDM model, is $\sim 10^{-1}\hmpc$. For massive neutrinos, this scale is typically $10^{-2}\hmpc$, while for allowed WDM models, it is $\gtrsim 1\hmpc$. Another point of difference from massive neutrinos is the fact that the small scale behavior is not completely degenerate with a change in $\sigma_8$. This has certain implications for the proposed role of LiMRs in resolving the $\sigma_8$ tension. Finally, the LiMR affects the mass-concentration relation of even the most massive halos, in contrast to viable WDM models which only affect the relationship for lower mass objects. 

It is worth noting that the particular LiMR model parameters that were explored in this work provide a good fit to the \textit{Planck} primary data. That is, early Universe observations by themselves are not sufficient to strongly constrain LiMR models. Given our findings, it is important to combine early Universe data with late time structure formation to obtain the best constraints on these models. Furthermore, including small, nonlinear scales of structure formation in such an analysis have the added benefit of being able to possibly distinguish between different velocity distributions of the LiMRs, even though they make the same predictions for the linear evolution on large scales.

   LiMR models have been proposed as a solution to the mild tension between early Universe and late time probes of clustering, parameterized in terms of $\sigma_8$. Note that SM neutrinos cannot, by themselves, account for the size of the discrepancy. Low redshift clustering probes that point to a $\sigma_8$ tension \cite[e.g.][]{2019PASJ...71...43H,2020JCAP...05..005D,2020Ivanov,2021A&A...649A..88T,2021arXiv210513549D,2021JCAP...12..028K}, do not include small nonlinear scales in their analysis. Therefore, one can always tune the parameters of a particular LiMR model to produce a value of late time $\sigma_8$ consistent with these studies. A full characterization of the nonlinear predictions of the models on small scales, as we have done here, is therefore crucial in making the models predictive and testable again. The fact that we find that the nonlinear effects of this particular LiMR model are not degenerate with an overall rescaling of the initial power spectrum within the $\Lambda$CDM framework implies that it is quite possible to rule out or constrain the model with available and future data. If the model is ruled out, other solutions to the $\sigma_8$ tension need to be pursued. 

While we have focused on a few of the signatures of LiMRs on small, nonlinear scales that are relevant for large-scale photometric surveys, their effects should also be seen on other observables. For example, the small scale damping in $P(k)$ will impact Lyman-$\alpha$ forest measurements, and therefore these measurements can be turned into effective constraints on LiMR models, as has been done for WDM \citep{2013PhRvD..88d3502V,2017PhRvD..96b3522I},  and Fuzzy Dark Matter (FDM) \citep{2017PhRvL.119c1302I,2021PhRvL.126g1302R} models. The damping also implies effects on the satellite population of the Milky Way, studies of which now provide some of the strongest constraints on various dark matter models \cite[see e.g.][]{2019ApJ...878L..32N,2021PhRvL.126i1101N,2021PhRvD.103d3517D}. We leave a thorough study of LiMR effects on these observables and possible constraints to a future work.

\section*{Acknowledgements}
The authors thank Chihway Chang and Eric Baxter for providing the DES Y1 measurements and error bars used in the paper. The authors thank Tom Abel, Yacine Ali-Ha{\"\i}moud, Steen Hannestad, Jessie Muir, Ethan Nadler and Francisco Villaescusa-Navarro for helpful comments on an earlier version of the manuscript. AB thanks Susmita Adhikari, Chihway Chang, and Josh Frieman for stimulating discussions. AM is supported
in part by the SERB, DST, Government of India by the grant MTR/2019/000267.SD acknowledges SERB grant CRG/2019/006147. AB is supported by the Fermi Research Alliance, LLC under Contract No. DE-AC02-07CH11359 with
the U.S. Department of Energy, and through the U.S. Department of Energy (DOE) Office of Science Distinguished Scientist Fellow Program. 

\pagebreak
\bibliography{sample631}{}
\bibliographystyle{aasjournal}

\appendix

\section{Effect of LiMR velocity distributions on nonlinear structure formation}
\label{sec:velocity_distribution}



\begin{figure}[htp!]
\centering
\includegraphics[scale=0.48]{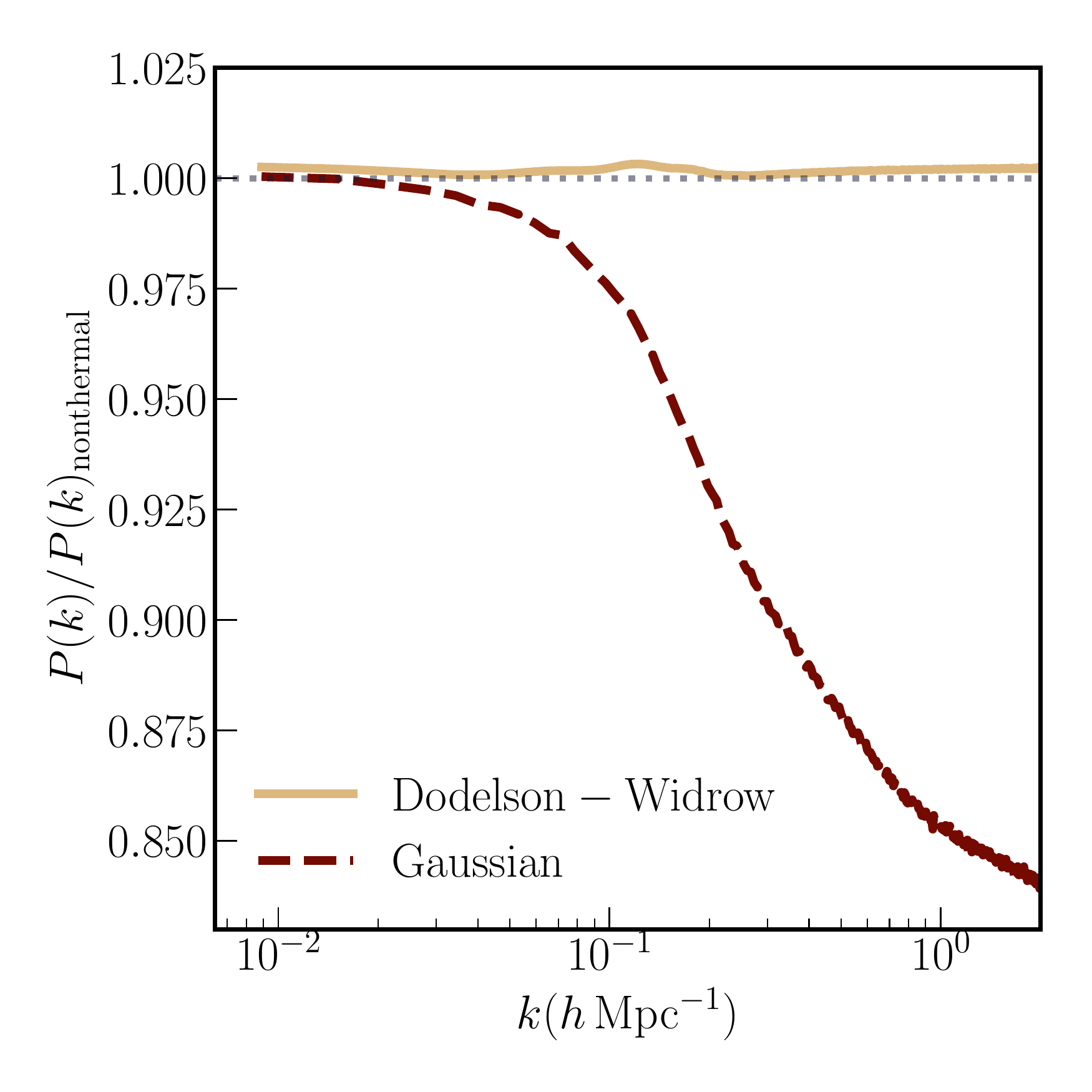}
\includegraphics[scale=0.48]{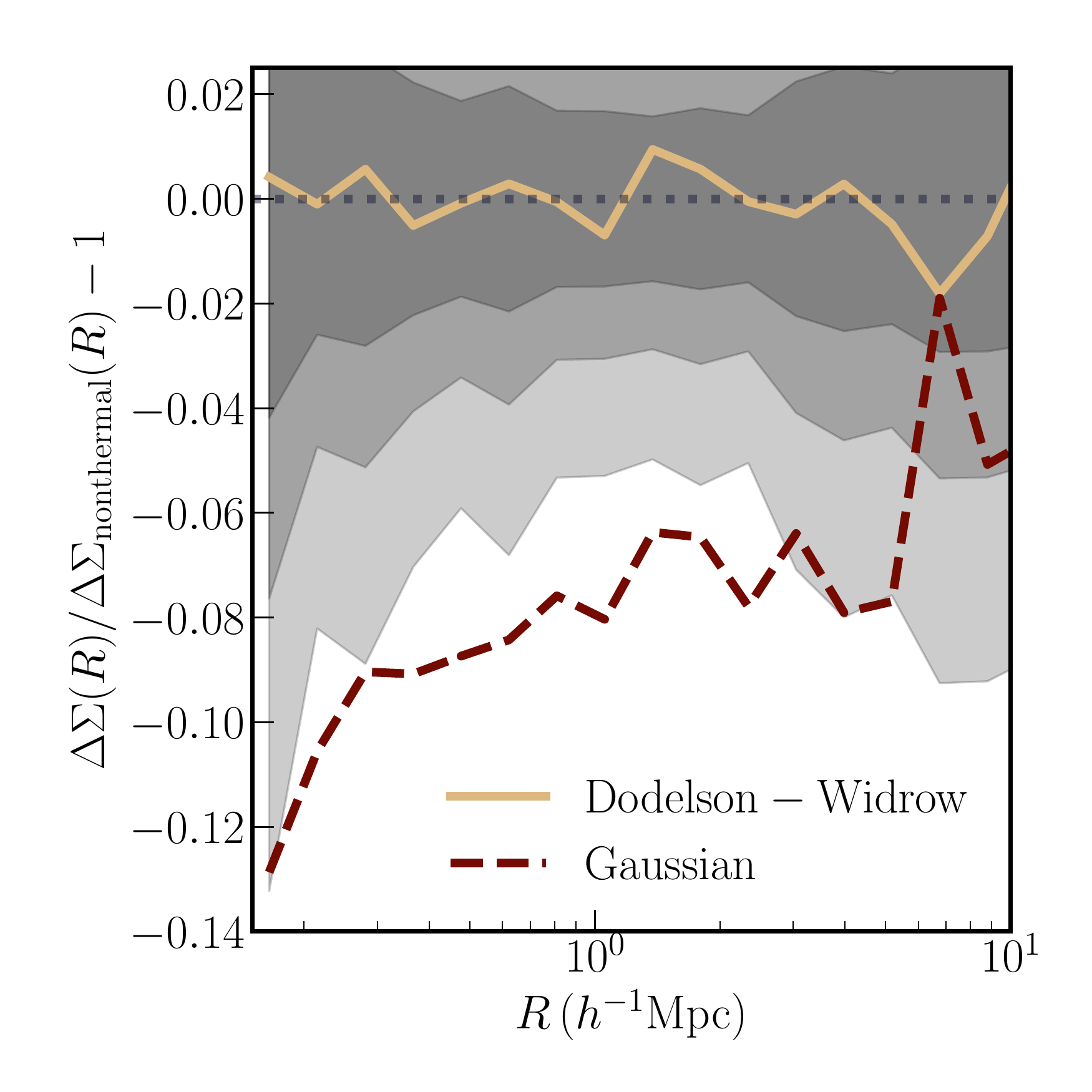}
\caption{\textit{Left panel:} Ratio of the total matter power spectrum at $z=0$ for the Dodelson-Widrow model (solid yellow curve) and the Gaussian model (dashed maroon curve), to the total matter power from the nonthermal LiMR model discussed in Sec. \ref{sec:results}. The dotted line is the expectation for no deviation from the fiducial nonthermal LiMR model. \textit{Right panel:} Residual of $\Delta \Sigma$ to the nonthermal LiMR model discussed in the main text. Once again, the solid yellow line is the residual for the Dodelson-Widrow model, and the dashed maroon line is the residual for the Gaussian model. The dotted line represents zero deviation from the fiducial nonthermal LiMR model. Shaded regions represent expected data error bars from DES-Y1, DES-Y3 and VRO LSST (see Sec. \ref{sec:cluster_lensing} for details on how the error bars are computed). Even with future surveys, it will be difficult to distinguish the fiducial nonthermal LiMR model and an LiMR model with a Dodelson-Widrow distribution. However, if the velocity distribution is sufficiently different, as is the case for the Gaussian distribution, the differences in the observables are large enough to be detected. Note that, in contrast to Figs.~\ref{fig:pk} and \ref{fig:delta_sigma}, the fiducial model in this figure (both panels) is the nonthermal model from  \citet{Das:2021pof}, rather than the \planck best-fit $\Lambda$CDM model.} 
\label{fig:pk_vel} 
\end{figure}

In this Appendix, we investigate the effect of changing the velocity distribution of the LiMR on the nonlinear signatures of the models at small scales ($\lesssim 10 \hmpc$). In this section, we will use the term ``fiducial" nonthermal LiMR model to refer to the one discussed in the main text of the paper \citep{Bhattacharya:2020zap}. We will explore the similarities and differences between the fiducial model, and the Dodelson-Widrow and Gaussian models discussed in Sec. \ref{sec:style}. The $N$-body simulations are carried out as described in Sec. \ref{sec:sims} --- the inputs that differ from model to model are the linear theory predictions from \textsc{CLASS} and the velocity distributions at $z=99$, as shown in Fig. \ref{fig:psd}. 
The halo finding is also carried out using the same methods discussed in Sec. \ref{sec:sims}.

The left panel of Fig. \ref{fig:pk_vel} illustrates the effect of changing the velocity distribution on the total matter power spectrum. The different curves represent the ratio of the power spectrum in that LiMR model to the power spectrum of the fiducial nonthermal LiMR model. The solid yellow line represents the ratio for the Dodelson-Widrow model, while the dashed maroon curve represents the ratio for the Gaussian model. The dotted line represents a ratio of exactly $1$ - i.e. no deviation from the fiducial model. The right panel of Fig. \ref{fig:pk_vel} shows the effects of a change in the LiMR velocity distribution on the weak lensing measurements, in terms of the excess surface density $\Delta \Sigma(R)$, around the $20000$ most massive halos in each simulation (see Sec. \ref{sec:cluster_lensing} for details of the simulation measurements). The residuals with respect to the fiducial nonthermal LiMR model are plotted. Once again, the yellow solid curve represents the residuals for the Dodelson-Widrow model, and the dashed maroon curve represents the residuals for the Gaussian model. The dotted line represents zero deviation from the fiducial model. The shaded grey regions represent the fractional error bars expected from measurements with sensitivity similar to DES Y1 (lightest grey band), DES Y3 and LSST (darkest grey band).

While interpreting the results, it is important to remember that the model parameters are tuned so that the linear theory predictions of all three LiMR models are the same. On small, nonlinear scales, we find that the results are no longer matched exactly, breaking the formal degeneracy. However, the difference between the fiducial nonthermal model and the Dodelson-Widrow model, remains quite small, and even with a survey as powerful as the LSST at VRO, there will not be sufficient sensitivity to distinguish between the two models. Note that this also implies that the results presented in Sec.~\ref{sec:results}, and the issues discussed in Sec.~\ref{sec:discussion}, while formally derived from the specific nonthermal LiMR model from \citet{Bhattacharya:2020zap}, can be generalized to the Dodelson-Widrow class of models. For the Gaussian model, which was tuned such that the velocity distribution is quite different from the other models, the small scale structure signatures are also quite distinct. In particular, the Gaussian model has very low occupancy at low velocities (see Fig. \ref{fig:psd}) compared to the other models. This low velocity tail is most relevant for nonlinear clustering, as shown in \citet{2018JCAP...09..028B} for massive neutrinos. Thus, the LiMR component in the Gaussian model clusters less, and produces a much larger damping in the power spectrum on small scales, as can be seen by comparing the yellow and maroon curves in the left panel of Fig. \ref{fig:pk_vel}. In terms of weak lensing measurements (right panel of Fig.~\ref{fig:pk_vel}) also, the Gaussian model is well separated from the other two for the same reason. In fact, the differences are large enough that it might be possible to distinguish such a model with a high statistical significance with future data. 

\end{document}